\documentclass[final,3p,times]{elsarticle}

\usepackage{latexsym,amssymb,epsfig}
\usepackage{amsmath,amssymb}
\usepackage{graphicx}
\usepackage{psfrag}

\usepackage{lineno}
\journal{Applied Numerical Mathematics}

\begin{document}

\begin{frontmatter}



\title{Pad\'e approximants and the prediction of non-perturbative parameters in particle physics}


\author{Oscar Cat\`a\fnref{label2}}
\ead{ocata@lnf.infn.it}
\fntext[label2]{Work supported by the EU under contract MTRN-CT-2006-035482 Flavianet.}
\address{Departament de F\'isica Te\`orica, IFIC, Universitat de Val\`encia – CSIC, Apt. Correus 22085, E-46071 Val\`encia, Spain}


\begin{abstract}
Commonly used techniques to study non-perturbative aspects of the strong interactions have a deep connection with rational approximants, and in particular with Pad\'e approximants to meromorphic functions. However, only recently this connection has been acknowledged and efforts at fully exploiting it are only starting. In this article I will briefly review the most prominent techniques used in non-perturbative strong interactions with special emphasis on its relation with Pad\'e approximants. I will then concentrate on a set of open problems outside the scope of these conventional techniques where Pad\'e approximants might be extremely useful. 


\end{abstract}


\end{frontmatter}




\section{Introduction}\label{sec1}
Particle physics is nowadays a mature discipline inside theoretical physics, where all observed phenomena among its elementary constituents (quarks and leptons) can be described in terms of four fundamental interactions: gravitational, strong, weak and electromagnetic. Unless one goes to extremely high energies, where gravitational effects are no longer negligible, strong, weak and electromagnetic interactions suffice to give a satisfactory description of the dynamics of elementary particles. Even though the nature of those three interactions is apparently very different, there is a common guiding principle, the gauge principle, that asserts that particle dynamics are determined by symmetries. 

In this article I will concentrate on the strong interactions, which describe the interactions of quarks mediated by gluons. Free quarks are described by the Dirac Lagrangian:
\begin{equation}
{\cal{L}}={\bar{\psi}}[i\gamma_{\mu}\partial^{\mu}-m]\psi~,
\end{equation}
while interactions are implemented by requiring invariance of the previous Lagrangian with respect to local $SU(3)$ transformations. This prompts the appearance of the gluon field $G_{\mu}$ and the final Lagrangian is
\begin{equation}\label{Lagr}
{\cal{L}}_{QCD}=-\frac{1}{2}{\mathrm{Tr}}\,G^{\mu\nu}G_{\mu\nu}+\sum_{q=1}^6{\bar{\psi}}_q[i\gamma_{\mu}D^{\mu}-m_q]\psi_q~,
\end{equation}
where $D^{\mu}=\partial^{\mu}-ig_sG^{\mu}$ and $[D_{\mu},D_{\nu}]=-ig_sG_{\mu\nu}$, $g_s$ being the strong coupling constant. The kinetic term is added to make the gluon a dynamical field. Thus symmetry fully determines the structure of the strong interactions, with the 6 quark species (flavours) belonging to the fundamental representation of $SU(3)$ and the gauge potential $G_{\mu}$ to the adjoint representation.

The simplicity of Eq.~(\ref{Lagr}) does however little justice to the complexities it hides. First of all, quarks and gluons have never been observed. Instead, experimentalists only detect hadrons, and a large number of them. Therefore, Eq.~(\ref{Lagr}) was initially received with scepticism and only taken as a model where hadrons were composite objects and quarks and gluons their building blocks. While this complied very well with some static properties of hadrons, there was no dynamical
explanation of this compositeness or {\emph{confinement}}: the equations of motion cannot be solved analytically, and perturbative expansions in the strong coupling $g_s$ seemed sheer nonsense.

A major breakthrough came with the realization that at extremely high energies $\mu$ the binding of quarks and gluons becomes loose, {\it{i.e.}}, 
\begin{equation}
\lim_{\mu \rightarrow \infty} g_s(\mu)=0~.
\end{equation}
This asymptotic freedom~\cite{Gross:1973id,Politzer:1973fx} of the strong interactions means that QCD at high energies behaves like a non-Abelian version of quantum electrodynamics (QED), and in particular makes perturbative expansions in the strong coupling (pQCD) meaningful. 

Parallel to the development of pQCD, techniques were developed to grasp some of the non-perturbative effects of QCD. The operator product expansion (OPE)~\cite{Wilson:1969zs,Shifman:1978bx} is a well-defined procedure to incorporate non-perturbative effects as inverse powers of momenta $q^2$. In practice, given a quantity $\Pi(q^2,g_s)$, at large values of $q^2$ one has a double expansion of the form
\begin{equation}
\Pi(q^2,g_s)=\sum_{n,m}\,g_s^n\log^m{q^2}\left(c_{nm}+\sum_k\frac{c_{nmk}}{q^{2k}}\right)~,
\end{equation}
where each of the series is believed to be asymptotic.

At very low energies, there is also (non-perturbative) information that can be extracted. The so-called chiral symmetry is broken and Goldstone theorem requires the presence of massless particles~\cite{Goldstone:1962es}, whose interactions are highly constrained by symmetry properties and where most of the parameters can be determined experimentally or estimated theoretically. Therefore, for every quantity in QCD there is accessible information at very high and very low Euclidean momenta. It therefore seems like an ideal setting to use rational approximants as interpolators to fill in the unknown region.

The applications of Pad\'e approximants to the study of the strong interactions is by no means a new topic. There was a big surge in the 1970's, in the pre-QCD times, where multiple models of hadrons existed. With the advent of QCD (and the realization that the theory was asymptotically free), perturbative computations took over those hadronic models. Nowadays not much remains of these pre-QCD applications in reference textbooks.   

New non-perturbative techniques were developed after QCD to extract information from known low and high energy input. However, their motivations did not come from rational approximants and, even though essentially all of the methods can be recast as Pad\'e approximants, the QCD community is only slowly acknowledging this close connection with Pad\'e theory.

In this article I will briefly review those non-perturbative standard methods, placing emphasis on their relation to Pad\'e approximants. Then I will describe in some detail a set of applications to open problems where more unconventional approaches based on Pad\'e theory have been attempted and where further work in that direction seems promising. Let me state from the beginning that this review is not intended to be comprehensive. It is rather biased towards those aspects of Pad\'e theory that I've come across in my own research. Likewise, the selection of applications is very personal.  


\section{The large-$N_c$ limit of the strong interactions}\label{sec2}
An obvious possibility to gain insight into the strong interactions is to play with the parameter space of the theory. However, the problem is that there is no obvious free parameter other than the quark masses. G.~'t~Hooft~\cite{'tHooft:1973jz} proposed in 1973 to consider the family of QCD-like theories with generic gauge group $SU(N_c)$. In the joined limit
\begin{equation}
N_c\rightarrow \infty,~\quad g^2N_c=ct.
\end{equation}
it can be shown that a consistent non-perturbative theory of hadrons emerges. The argument is very simple. Let us consider, for concreteness, a generic two-point correlator with arbitrary sources. Its full result in QCD can be expressed (formally at least) as a series expansion in the strong coupling constant. Whether this infinite series is summable is not important at this point. The important point is that the full perturbative expansion has to be dual to a hadronic realization in order to preserve unitarity. 

\begin{figure}[h!]
\centering
\includegraphics[width=4.5in]{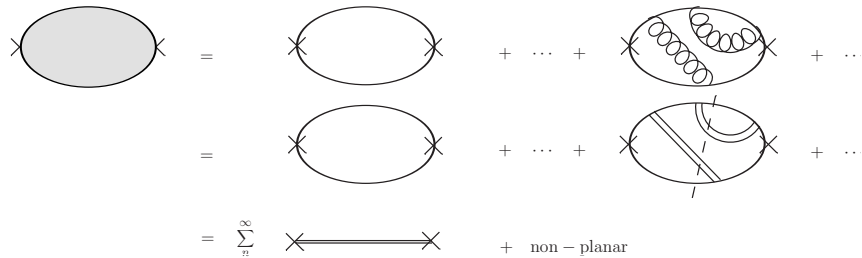}
\caption{Pictorical representation of a generic two-point correlator (left-hand side) in terms of its Feynman diagram perturbative series expansion in both conventional (first line) and double-line (second line) notations, together with its dual hadronic representation in terms of the topological expansion (last line). The generic sources are represented by crosses.}\label{fig1}
\end{figure}

It is standard to represent the perturbative series pictorically as Feynman diagrams, some of which are shown in the first line of Fig.~\ref{fig1}. The argument can be best understood by introducing the double-line notation, as shown in the second line of Fig.~\ref{fig1}, where gluon lines are depicted as double lines (because they belong to the adjoint representation of $SU(N_c)$) and quarks as single lines (because they belong to the fundamental representation). 
Double-line notation thus offers an easy way to keep track of the colour flow inside a given diagram.\footnote{group indices in $SU(N_c)$ are normally referred to as colour indices.} If confinement is assumed, then no open colour lines can exist. With the double-line notation it is not difficult to see that the $1/N_c$ expansion is actually a topological expansion, where planar diagrams dominate. Without solving the full set of planar diagrams, the optical theorem allows to connect the quark-gluon and the hadronic picture: the dashed line crossing the two-gluon correction in the second line of Fig.~\ref{fig1} gives a single colour-singlet ${\bar{q}}q$ state, {\it{i.e.}} a single meson exchange. In fact, any possible way of cutting any planar diagram will lead to the same conclusion~\cite{Witten:1979kh}. Always at a qualitative level, one can show that the hadronic picture dual to planar diagrams consists of the exchange of an infinite number of single and stable resonances. In a more mathematical language, this implies that the correlator is meromorphic, consisting of an infinite number of single poles. Therefore for the generic two-point correlator $\Pi(q^2)$, its spectral function reads
\begin{equation}
\frac{1}{\pi}\,{\mathrm{Im}}\,\Pi(t)=\sum_n^{\infty}f_n^2\delta(t-m_n^2)~, 
\end{equation}
where $m_n^2$ are simple poles and $f_n^2$ their associated residues. This spectral representation can be related to the full correlator by using Cauchy's theorem on the Argand diagram of Fig.~\ref{fig3}, which shows the singularity structure of a generic two-point correlator. Therefore
\begin{equation}\label{lar}
\Pi(q^2)=\int_0^{\infty}\frac{dt}{-q^2+t}\frac{1}{\pi}\,{\mathrm{Im}}\,\Pi(t)+{\cal{A}}(q^2)=\sum_n^{\infty}\frac{f_n^2}{-q^2+m_{n}^2}+{\cal{A}}(q^2)~,
\end{equation}
where ${\cal{A}}(q^2)$ is a polynomial that depends on the convergence properties of $\Pi(q^2)$ at infinity.

Unfortunately, an analytical solution of large-$N_c$ QCD is still lacking. However, even being qualitative, the meromorphic representation of Eq.~(\ref{lar}) for the hadronic spectrum is currently the best available description of the hadronic world, and is at the base of almost all the phenomenological methods developed so far to study non-perturbative QCD. 

\section{Non-perturbative methods as Pad\'e approximants}\label{sec3}
As already emphasized in the introduction, many of the non-perturbative methods in strong interaction physics developed after the advent of QCD are, sometimes in a disguised way, realizations of Pad\'e approximants. The underlying reason is that all the methods are based, to a certain degree, on meromorphic ans\"atze subject to some matching to low and/or high energy QCD inputs. In this section I will review some of the most well-established ones. 

As an illustrative example for all the methods, I will consider the following correlators:
\begin{eqnarray}
\Pi^{V}_{\mu\nu}(q)&=&i\int\mathrm{d}^{4}x\, e^{iq\cdot x}\langle \,0\,|\,T\lbrace\, V_{\mu}(x)\,V_{\nu}^{\dagger}(0)\,\rbrace |\,0\,\rangle=(q^{\mu}q^{\nu}-q^2g^{\mu\nu})\,\Pi_{V}(q^2)~,\label{correl1}\\
\Pi^{A}_{\mu\nu}(q)&=&i\int\mathrm{d}^{4}x\, e^{iq\cdot x}\langle \,0\,|\,T\lbrace\, A_{\mu}(x)\,A_{\nu}^{\dagger}(0)\,\rbrace |\,0\,\rangle=(q^{\mu}q^{\nu}-q^2g^{\mu\nu})\,\Pi_{A}(q^2)~,\label{correl2}
\end{eqnarray}
where $V_{\mu}(x)=\bar{u}(x)\gamma_{\mu}d(x)$ and $A_{\mu}(x)=\bar{u}(x)\gamma_{\mu}\gamma_5 d(x)$ are QCD currents and the factorization of the tensor structure follows from Lorentz and gauge symmetries. Each of the previous correlators diverges logarithmically at large values of $q^2$, where perturbation theory can be used. However, the coefficient in front of the logarithm is the same for the vector and axial channel, which is a consequence of chiral symmetry being unbroken at the quark-gluon level. Thus, the difference $\Pi_{VV}-\Pi_{AA}\equiv\Pi_{LR}$ is only sensitive to non-perturbative physics and in particular to the breaking of chiral symmetry at the hadronic level. One normally denotes such quantities as {\emph{order parameters}} of chiral symmetry breaking. Therefore, at high momenta the correlator can be parameterized entirely by the OPE as
\begin{equation}\label{opevector}
\lim_{q^2\rightarrow (-\infty)}\Pi_{LR}(q^2)=\sum_{n=3}^{\infty}\frac{\xi_{2n}}{q^{2n}}~.
\end{equation}  
It can be shown that for this particular correlator the first two terms, $\xi_2$ and $\xi_4$, cancel to a very good approximation.
 
As a common background, all the methods to be described adopt the large-$N_c$ limit version of QCD described in the previous section. Therefore, the spectral function for $\Pi_{LR}$ can be written as
\begin{equation}
\frac{1}{\pi}\,{\mathrm{Im}}\,\Pi_{LR}(t)=-f_{\pi}^2\delta(t)+\sum_n^{\infty}f_{Vn}^2\delta(t-m_{Vn}^2)-\sum_n^{\infty}f_{An}^2\delta(t-m_{An}^2)~,
\end{equation}
and the correlator as{\footnote{Due to the super-convergence of $\Pi_{LR}$ at high energies, the function ${\cal{A}}(q^2)$ actually vanishes.}
\begin{equation}
\Pi_{LR}(q^2)=\frac{f_{\pi}^2}{q^2}+\sum_n^{\infty}\frac{f_{Vn}^2}{-q^2+m_{Vn}^2}-\sum_n^{\infty}\frac{f_{An}^2}{-q^2+m_{An}^2}~,
\end{equation}
where the presence of the massless pion mode is a direct consequence of Goldstone's theorem.
 
The differences between the different methods to be described come from the approximations made on the previous equation, the degree of matching to the underlying information from the fundamental QCD theory, and the different non-perturbative quantities one wants to compute. 

\subsection{Vector meson dominance}

Vector meson dominance (VMD)~\cite{Sakurai} rests on the hypothesis that whenever vector mesons are allowed by quantum numbers to contribute to a certain process, then (i) they will dominate over other particle channels and (ii) their lowest-lying states (the poles closer to the origin) will give the bulk of the non-perturbative effects. 

Therefore, the spectral function of Eq.~(\ref{lar}) will be completely saturated by a single resonance on each channel, namely 
\begin{equation}
\Pi_{LR}(q^2)=\frac{f_{\pi}^2}{q^2}+\frac{f_V^2}{-q^2+m_V^2}-\frac{f_A^2}{-q^2+m_A^2}~.
\end{equation}
An additional key ingredient of VMD is that the poles and residues of the previous equation are identified with the physical parameters of the lowest-lying states in the hadronic spectrum, {\it{i.e.}} the $\rho(770)$ and $a_1(1260)$ mesons. Therefore, in the ansatz above, $m_V\equiv m_{\rho}=770$ MeV, $m_A\equiv m_{a1}=1260$ MeV, $f_V\equiv f_{\rho}\simeq 218$ MeV and $f_A\equiv f_{a1}\simeq 174$ MeV~\cite{Ecker:1988te}.

\subsection{Sum rules: Finite energy and Borel-Laplace sum rules}

An important problem with the VMD approach is that it requires precise knowledge of physical parameters, which typically is scarce, especially for the residues. The sum rule approach~\cite{Shifman:1978bx} overcomes this drawback, because it computes residues and/or poles in terms of a few universal OPE terms.

The physical picture behind the sum rules is that the hadronic world, represented by the ans\"atze for the spectral function, should match QCD at high energies. Recall that in order to carry out this program one needs in general an infinite number of resonances in the spectral function, otherwise the logarithms of pQCD cannot be reproduced. In practice, and in order to be predictive, the spectral ansatz for a generic correlator $\Pi(q^2)$ consists of simple poles at $m_n^2$ and a cut starting at a certain threshold $s_0$, {\it{i.e.}},
\begin{equation}
{\mathrm{Im}}\,\Pi(t)=\sum_n^{N}f_{n}^2\delta(t-m_{n}^2)+A\theta(t-s_0)~.
\end{equation} 
Notice that despite being more predictive, this modeling of the continuum as a Heavyside-$\theta$ function departs from the large-$N_c$ version of an infinite set of Dirac-$\delta$ functions.

There are clear advantages when one uses sum rules on order parameters. First and foremost, since perturbation theory cancels to all orders, one expects to be more sensitive to non-perturbative effects. Note also that in principle one could dispense with the continuum contribution without conflicting with the OPE.

For $\Pi_{LR}$, as already mentioned, not only perturbation theory cancels, but also the terms $q^{-d}$ with $d\leq 6$ in the OPE. This means that the matching equations read
\begin{equation}
\begin{array}{ccccccc}
\xi_2&=&-\int_0^{\infty}dt {\mathrm{Im}}\,\Pi_{LR}(t)&=&\displaystyle\sum_n^{{\cal{N}}_A}f_{An}^2-\sum_{n}^{{\cal{N}}_V}f_{Vn}^2+f_{\pi}^2&\equiv& 0\nonumber\\
\xi_4&=&-\int_0^{\infty}dt\,t\, {\mathrm{Im}}\,\Pi_{LR}(t)&=&\displaystyle\sum_n^{{\cal{N}}_A}f_{An}^2m_{An}^2-\sum_n^{{\cal{N}}_V}f_{Vn}^2m_{Vn}^2&\equiv& 0\nonumber\\
\vdots&=&\vdots&=&\vdots&&\nonumber\\
\xi_{2j}&=&-\int_0^{\infty}dt\,t^{j-1}\, {\mathrm{Im}}\,\Pi_{LR}(t)&=&\displaystyle\sum_n^{{\cal{N}}_A}f_{An}^2m_{An}^{2j-2}-\sum_n^{{\cal{N}}_V}f_{Vn}^2m_{Vn}^{2j-2}~,&&
\end{array}
\end{equation}  
where $\xi_{2j}$ are the OPE coefficients defined in Eq.~(\ref{opevector}). In the sum rule approach the poles are taken as input from the experimental masses of the lowest-lying states in the spectrum. With $\Pi_{LR}$, experience indicates that the matching equations above can be satisfied with the minimal hadronic content. However, in certain cases this might turn out to be a bad approximation and contributions from higher order poles can be important. Instead of adding those contributions explicitly, a common strategy is to use Borel sum rules.

With Borel sum rules, one starts from
\begin{equation}
\int_0^{s_0}dt\, e^{-t\tau}\,\frac{1}{\pi}\,{\mathrm{Im}}\,\Pi_{LR}(t)=-\frac{1}{2\pi i}\oint_{|q^2|=s_0}\, e^{-q^2\tau}\,\Pi_{LR}(q^2)~,
\end{equation}
which again follows from Cauchy's theorem applied to Fig.~\ref{fig3}, where $s_0$ is the radius of the circle. On the left-hand side one plugs in the hadronic ansatz while on the right-hand side one uses the OPE.\footnote{Strictly speaking, plugging the OPE is inconsistent with Cauchy's theorem. In general, on the right-hand side one should also include terms that account for quark-hadron duality violation. I will not delve into this issue and instead refer the reader to~\cite{Cata:2005zj}, where a systematic study of such effects on the $\Pi_{LR}$ correlator is performed.} In general one will find
\begin{equation}
\sum_n^{N_V}f_{Vn}^2 e^{-m_{Vn}^2\tau}-\sum_n^{N_A}f_{An}^2 e^{-m_{An}^2\tau}-f_{\pi}^2=\sum_j^{N}\frac{(-1)^j}{(d-1)!}\xi_{2j} \tau^{j-1}
\end{equation}
Notice that higher values of the Borel parameter $\tau$ suppress exponentially the contributions from heavier states on the hadronic side, but at the same time enhance the weight of higher dimension operators in the OPE side. In practice one should look for some stability ({\it{i.e.}} $\tau$-independent window), in the Borel plane. 

\subsection{The minimal hadronic approximation}\label{sec32}

The minimal hadronic approximation (MHA)~\cite{Peris:1998nj,Knecht:1999kg} was initially devised to compute non-perturbative quantities that can be expressed as integrals over the Euclidean regime of correlators. One such instance is the pion electromagnetic mass difference, given by~\cite{Das:1967it} 
\begin{equation}
\Delta m_{\pi}=-\frac{3\alpha}{4\pi f_{\pi}^2}\int_0^{\infty}dQ^2Q^2\Pi_{LR}(Q^2)~,\qquad (Q^2=-q^2)~.
\end{equation}
The MHA assumes the correlator to be meromorphic, with the minimal number of poles to reproduce the OPE behavior known from QCD. It rests on three key points:
\begin{itemize}
\item the importance of matching to both low and high energies, {\it{i.e.}}, to the Laurent (OPE) and Taylor (chiral) expansions of the correlator under study. There are different reasons to prefer this 2-point matching to a simple 1-point one: first, the more information one can implement in the ansatz the better, and the first terms in both the Taylor and Laurent expansions are known with reasonable accuracy. Second, requiring matching on both high and low energies improves the stability of the result, especially since one is interested not in the local behavior of $\Pi_{LR}$ but on the area below it. See for instance Fig.~\ref{fig2} borrowed from Ref.~\cite{DeRafael:2001zs}.
\item The poles of the meromorphic function are identified with physical particle masses, while the residues are determined from the matching equations.
\item The method is restricted to correlators that are order parameters, {\it{i.e.}}, those for which the contribution from pQCD identically vanishes. In other words, the meromorphic ansatz is only compatible with ultraviolet finite correlators, namely those that behave like inverse powers of $q^2$.\footnote{Even in this case the method has its limitations. For instance, it has been realized~\cite{Bijnens:2003rc} that a meromorphic ansatz with a finite number of poles cannot fulfill all the high energy constraints, especially those coming from quark-counting rules.} Correlators with Euclidean logarithms are therefore beyond the scope of the method. In practice however this is not a severe restriction on the number of observables to compute.  
\end{itemize}
\begin{figure}
\begin{center}
\includegraphics[width=3.8 in]{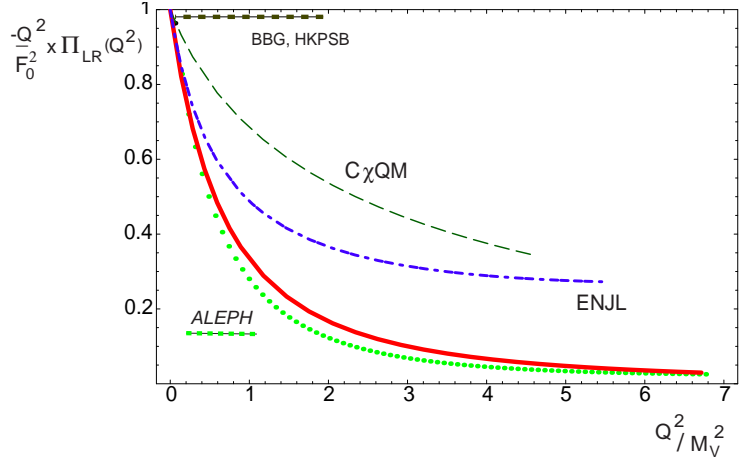}
\caption{Comparison of the MHA ansatz (solid red line) with experiment (dotted green line) on $\Pi_{LR}$ together with the predictions of other methods without high energy matching. Figure borrowed from Ref.~\cite{DeRafael:2001zs}.}\label{fig2}
\end{center}
\end{figure}
For further details and different applications of the MHA to electroweak physics I refer the interested reader to Refs.~\cite{Cata:2003mn,Cata:2004ti,Hambye:2003cy,Knecht:2002hr,Knecht:2001bc,Peris:1998nj,Peris:2000sw}.   

From the previous discussion it is evident that both sum rules and the MHA are straightforward applications of Pad\'e-type approximants. FESR are 1-point Pad\'e-type approximants, BSR are 1-point Borel-Pad\'e-type approximants while the MHA is an application of 2-point Pad\'e-type approximants.  Nonetheless, let me emphasize once again that all the previous methods did originate without prior knowledge of Pad\'e theory. Actually, in all of them, the parameters appearing in the ansatz (poles and residues) are considered as physical parameters, something unjustified from the Pad\'e theory perspective, where the ansatz is seen as a mere interpolator. I am fully convinced that these prejudices behind the particle interpretation of the ansatz are precisely what has prevented to move from Pad\'e-type to regular Pad\'e approximants in non-perturbative QCD. After all, it is well known that Pad\'e poles are not restricted to be real-positive. In any case, the fact that the previous approaches turned out to be realizations of Pad\'e approximants is not only remarkable {\emph{per se}}, but at the same time makes Pad\'e approximants compelling as natural tools to address non-perturbative QCD problems. 

\section{Some applications}\label{sec4}
In this section I will describe in some detail three different open problems in non-perturbative QCD that cannot be addressed with the techniques reviewed in the last section. However, the potential of Pad\'e approximants to make progress in each of them is far from being exhausted.  
\subsection{Pad\'e approximants and the hadronic spectrum}\label{sec41}
At the end of the 1970's there was an attempt to determine the spectrum of QCD (or at least the one of large-$N_c$ QCD) using Pad\'e approximants. To the best of my knowledge, this is one of the first applications of Pad\'e theory to non-perturbative particle physics after the advent of the QCD Lagrangian. In the following I will briefly outline the original derivation of Ref.~\cite{Migdal:1977nu}. 

The Pad\'e approximant to $\Pi_V(q^2)$ ({\it{cf.}} Eq.~(\ref{correl1})) around a point is given by the following equation:  
\begin{equation}
\Pi_V(q^2)= \Pi_V^{[N,M]}(q^2)+{\cal{R}}_{[N,M]}(q^2)\, , \qquad  \Pi_V^{[N,M]}(q^2)\equiv \frac{{\cal{P}}_M(q^2)}{{\cal{Q}}_N(q^2)}~,
\end{equation}
where by construction $\Pi_V^{[N,M]}(q^2)$ reproduces the first $N+M+1$ derivatives of $\Pi_V(q^2)$ around the point. In the following I will consider the diagonal sequence $[N,N]$. Hence it follows that
\begin{equation}\label{sec} 
\frac{d^n}{d(q^2)^n}\bigg[\Pi_V(q^2){\cal{Q}}_N(q^2)-{\cal{P}}_N(q^2)\bigg]\bigg|_{q^2=-\mu^2}=0,\qquad n=0,...,2N~.
\end{equation}
By construction, ${\cal{P}}_N$ cancels after the first $N$ equations, and applying Cauchy's theorem one ends up with 
\begin{equation}\label{pad}
\int_0^{\infty}\frac{dt}{(t+\mu^2)^{n+1}}\,{\mathrm{Im}}\Pi_V(t){\cal{Q}}_N(t)=0,\qquad n=N+1, ... , 2N~,
\end{equation}
from which one can determine the denominator ${\cal{Q}}_N$. So far no approximation has been made. In order to proceed further and solve the previous equation some information has to be provided on ${\mathrm{Im}}\Pi_V(t)$. Back in 1977 very little was known about non-perturbative expansions (the operator product expansion is from 1978), so the natural starting point was the perturbative expansion, whose leading term reads
\begin{figure}[t]
\centering 
\psfrag{A}{${\mathrm{Re}}\, q^2$}
\psfrag{B}{$\gamma$}
\psfrag{C}{\footnotesize{$(q^2=-\mu^2)$}}
\includegraphics[width=2.0in]{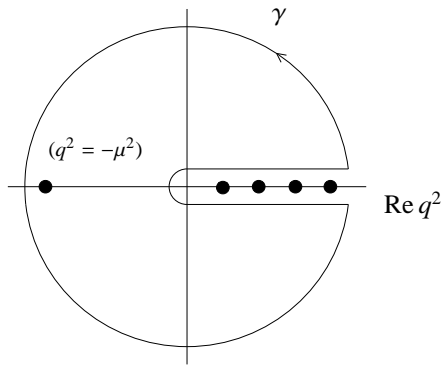}
\caption{Singularity structure of two-point correlators $\Pi(q^2)$ in the complex $q^2$ plane with the contour chosen to apply Cauchy's theorem. The circle is defined by $|q^2|=s_0$, which will be eventually sent to infinity. The singularities on the right half-plane are the poles of $\Pi(q^2)$, sitting at $m_n^2$. The point on the left half-plane is the pole with multiplicity $(n+1)$ explicitly shown in the denominator of Eq.~(\ref{pad}).}\label{fig3}
\end{figure}
\begin{equation}
\lim_{q^2\rightarrow (-\infty)}\Pi_V(q^2)=\frac{N_c}{12\pi^2}\log\left(\frac{-q^2}{\mu^2}\right)+\cdots
\end{equation}
This selects $q^2=-\mu^2$ as a natural expansion point, depicted in Fig.~\ref{fig3}. With the previous input, ${\mathrm{Im}}\Pi_V(t)$ is simply a constant and Eq.~(\ref{pad}) turns out to have a simple solution, namely 
\begin{equation}\label{result}
{\cal{Q}}_N(q^2)=_{\,\,2}\!\!F_1\left(-N,-N;1;-\frac{q^2}{\mu^2}\right)=(q^2+\mu^2)^N\,P_N^{(0,0)}\left(\frac{\mu^2-q^2}{\mu^2+q^2}\right)\, ,
\end{equation}
where ${}_{\,2}F_1(a,b;c;d)$ is Gauss' hypergeometric function and $P_N^{(0,0)}$ stands for the $(0,0)$-Jacobi polynomials, {\it{i.e.}}, the Legendre polynomials. Plugging the solution back in Eq.~(\ref{sec}) one can determine ${\cal{P}}_N(q^2)$. The result for the Pad\'e approximant eventually reads~\cite{Weideman}
\begin{equation}
\Pi_V^N(q^2)\simeq \frac{2}{(q^2+\mu^2)^N}\,\sum_{k=0}^{N}\left(\begin{array}{c}
k\\
j
\end{array}\right)^2 \left[\frac{H_{N-k}-H_k}{P_N^{(0,0)}(\chi)}\right]\left(-\frac{q^2}{\mu^2}\right)^k~; \qquad \chi=\frac{\mu^2-q^2}{\mu^2+q^2}~.
\end{equation}
The problem therefore has reduced to computing the Pad\'e approximant to the natural logarithm, a result originally found by Jacobi, Rouch\'e and Gauss back in the 19th century. 

Convergence of the Pad\'e approximant means that the original function is recovered when one lets $N\rightarrow \infty$. In the case of the logarithm it is well known that the set of Pad\'e poles becomes dense over the physical axis eventually mimicking the logarithmic cut. However, the main motivation behind Ref.~\cite{Migdal:1977nu} was to extract a set of poles from a continuum and a Pad\'e clearly does not serve this purpose. The following correlated limit was taken instead,
\begin{equation} 
q^2<<\mu^2,\quad N\rightarrow \infty,\quad \frac{N}{\mu}=ct.~,
\end{equation}
leading to the following difference in the resulting functions:
\begin{eqnarray}
\Pi_{V}^N(q^2)&\rightarrow& -\frac{4}{3}\frac{N_c}{(4\pi)^2}\log\frac{-q^2}{\mu^2}-\frac{{\cal{R}}_N(q^2)}{{\cal{Q}}_N(q^2)} \qquad \Big\{N\rightarrow \infty\Big\}\label{first}~,\\
\Pi_{V}^N(q^2)&\rightarrow& -\frac{4}{3}\frac{N_c}{(4\pi)^2}\left[\log\frac{q^2}{\mu^2}-\pi\frac{Y_0\left(qy_m\,\,\right)}{J_0\left(qy_m\,\,\right)}\right]\,\,\quad \Big\{N\rightarrow \infty,\,\, q^2<<\mu^2,\,\, \frac{N}{\mu}=ct.\Big\}~.\label{second}
\end{eqnarray}
Notice that even after the continuum limit is taken, in the second line there are a set of poles sitting at the zeroes of the Bessel $J_0$ function (see Ref.~\cite{Cata:2006ak} for a discussion on the physical interpretation of this last term). Those poles were claimed to give the spectrum of $\Pi_V(q^2)$, in other words, the spectrum of vector mesons.

There are a set of criticisms one can make to this approach, the most obvious being that the whole procedure is not a Pad\'e approximant, contrary to what it was initially advertised. On the physics side, some comments are in order too. First and foremost, I find it difficult to justify that starting from the perturbative quark-gluon logarithm one can infer the spectrum of hadrons, a purely non-perturbative entity. Second, but closely related, constructing the Pad\'e around an asymptotically large point $\mu^2$ is problematic: the expansion at large momenta is an asymptotic expansion of the original function, hence it does not uniquely define the original function.  In other words, the resulting meromorphization of the function is not unique. In fact, one can easily show that different spectra can reproduce at high energies the same partonic logarithm~\cite{Cata:2006ak}. This would be avoided if the Pad\'e is constructed around a finite point, for instance $q^2=0$~\cite{Peris:2006ds}, where data exists and the chiral expansion is well defined.
 
Although the solution proposed by Migdal turns out to be wrong for several reasons, the problem is very interesting and its solution not yet settled. I honestly believe that a proper application of Pad\'e approximants can improve the current situation. 

For instance, one could start by feeding Eq.~(\ref{pad}) with OPE condensates, which would certainly incorporate non-perturbative physics into the problem. In the same spirit, one could argue that some (Borel) resummation of the perturbative series might also help: to the extent that the asymptotic character of the perturbative expansion is thought to be a consequence of the existence of non-trivial infrared physics, there is definitely non-perturbative information lurking behind pQCD. This is at the base of the theory of renormalons (see the contribution of Prof. J.~Fischer in this conference). 

While I cannot exclude that this approach can yield some benefits, I foresee two potential difficulties: first, the perturbative expansion is plagued with logarithms, which cannot be used as input in a conventional Pad\'e approximant; and second, even if the previous difficulty is circumvented, one has to make sure that the Borel resummation captures a substantial amount of non-perturbative effects. 

What seems more feasible to me is to start from an expansion around the origin, in the spirit of~\cite{Peris:2006ds}. Then no logarithms are present, and the non-perturbative physics can be easily implemented from the following moments:
\begin{equation}\label{moments}
\int_0^{\infty}\frac{dt}{t^n}\,{\mathrm{Im}}\,\Pi_{V}(t)= \zeta_n~,
\end{equation}
where $\zeta_n$ are the coefficients of the MacLaurin expansion of $\Pi_{V}$, which can be extracted from experiment. Additionally, it can be shown that $\Pi_V$ is a function of Stieltjes type~\cite{Peris:2006ds}. Therefore, the Pad\'e poles all lie on the physical axis and convergence of the Pad\'e approximant (at least for finite $q^2$ and out of the physical axis) is guaranteed.

I do not know how far this approach can go. Certainly there are theorems like Koenig's theorem~\cite{Koenig} to approximate the poles of a function. The problem in QCD is that the number of poles is infinite, and I do not know of any result that applies in this case. 
\subsection{Pad\'e approximants and the gauge/string duality}\label{sec42}

Quite recently there have been attempts to infer the spectrum of QCD from theories defined in $(4+d)$ dimensions. The idea that our space-time, {\it{i.e.}}, a 4-dimensional manifold with metric $\eta_{\mu\nu}={\mathrm{diag}}(1,-1,-1,-1)$, and the quantum fields defined in it could be dynamically generated from a higher dimensional manifold is quite old. More recently, this kind of constructions have been suggested by superstring theories, where the cancellation of anomalies requires theories to be constructed in ten or eleven dimensions. Since in our everyday life we only perceive 4 dimensions (3 space-like and 1 time-like), the remaining extra dimensions have to be dynamically compactified. What has made those ideas more compelling are the recent developments following the so-called AdS/CFT correspondence (see for instance Ref.~\cite{Aharony:1999ti} and references therein), that gives a precise and well-defined prescription to connect a 5-dimensional manifold endowed with Anti-de Sitter (AdS) metric 
\begin{equation}
ds^2=g_{MN}dx^Mdx^N=\frac{1}{y^2}(-dy^2+\eta_{\mu\nu}dx^{\mu}dx^{\nu})~,
\end{equation}
to a 4-dimensional submanifold, through the following identity~\cite{Witten:1998qj}:
\begin{equation}
S_{5D}[\phi_0]=\int d^5x\, {\cal{L}}_{5D}(\phi_0)=\int d^4x\, ({\cal{L}}_{4D}+\phi_0 O_i)=S_{4D}+{\mathrm{sources}}~,
\end{equation}
where $\phi_0$ are the solutions of the 5-dimensional equations of motion projected on the 4-dimensional manifold and $O_i$ are operators in the 4-dimensional theory. Therefore, the correspondence claims that the 5-dimensional field solutions on the 4-dimensional submanifold are the sources of the correlators in the 4-dimensional theory. Even in these rather exotic scenarios Pad\'e approximants might help.
\begin{figure}[t]
\begin{center}
\includegraphics[width=8.0cm]{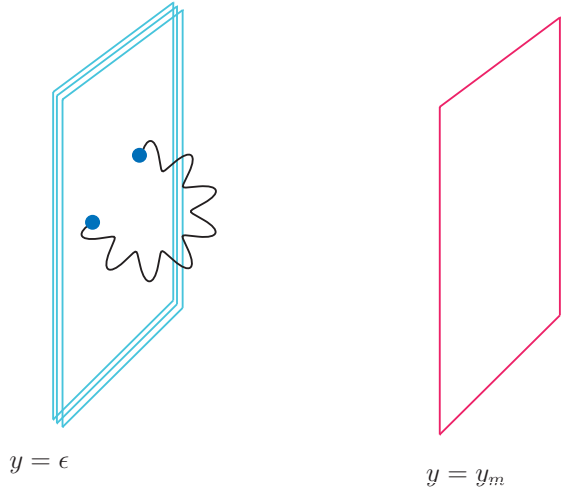}
\caption{Geometric setting used in Section~\ref{sec42}. The left-hand side manifold is the UV brane, where the gauge theory is defined. The right-hand side brane (IR brane) defines a particular model of confinement. The space in between, also called the bulk, is a 5-dimensional manifold with Anti-de Sitter metric.}\label{fig4}
\end{center}
\end{figure}  

Let us concentrate, as before, on vector mesons. We start from the massless Yang-Mills action in five dimensions:
\begin{equation}
S=\frac{1}{2g_5^2}\int d^4x\int_{\epsilon}^{y_m} dy \sqrt{g}\,{\mathrm{Tr}} \big[F_{MN}F^{MN}\big]~,
\end{equation}
where the 4-dimensional boundary branes are located at $y=\epsilon\rightarrow 0$ (high-energy or ultraviolet brane) and $y=y_m$ (low-energy or infrared brane). The setting is depicted in Fig.~\ref{fig4}. Without loss of generality one can set $V_5=0$. Then the equation of motion for the $V_{\mu}$ field reads
\begin{equation}
\left(\partial_y^2-\frac{1}{y}\partial_y-\Box\right)V_{\mu}=0~,
\end{equation}
which, subject to the following boundary conditions
\begin{equation}
\partial_y V_{\mu}(y_m)=0;\qquad V_{\mu}(\epsilon)=V_{\mu}^{(0)}~,
\end{equation}
results in~\cite{Pomarol:1999ad} 
\begin{equation}
{\hat{V}}_{\mu}(q,y)=\frac{y}{\epsilon}\frac{Y_0(qy_m)J_1(qy)-J_0(qy_m)Y_1(qy)}{Y_0(qy_m)J_1(q\epsilon)-J_0(qy_m)Y_1(q\epsilon)}V_{\mu}^{(0)}(q)\equiv {\hat{V}}(q,y) V_{\mu}^{(0)}(q)~.
\end{equation}
Following the AdS/CFT prescription, the previous solution can be plugged back into the action, leaving only a boundary term
\begin{equation}
S=-\frac{1}{2g_5^2}\int d^4x\,\frac{1}{y}{\hat{V}}_{\mu}\partial_y {\hat{V}}^{\mu}\Big|_{y=\epsilon\rightarrow 0}~.
\end{equation}
Notice that $V_{\mu}^{(0)}$, which is left unspecified, is the value of the field $V_{\mu}$ at the UV brane, and therefore also the source of correlators in the 4-dimensional theory. Therefore, $\Pi_V$ is given by~\cite{Erlich:2005qh}
\begin{eqnarray}
\Pi_{V}^{\mu\nu}&=&\frac{i\delta}{\delta V^{(0)}_{\mu}}\frac{i\delta}{\delta V^{(0)}_{\nu}}S\nonumber\\
&=&(q^{\mu}q^{\nu}-q^2g^{\mu\nu})\left[\frac{1}{g_5^2 q^2}\frac{1}{y}\partial_y{\hat{V}}\right]_{y=\epsilon}~.
\end{eqnarray}
The solution of the previous equation is exactly the one given by Eq.~(\ref{second}) in the previous section, with the identification
\begin{equation}
y_m=\frac{2N}{\mu}~.
\end{equation}
In other words, Migdal's prescription and the holographic model described before turn out to describe the same 4-dimensional theory. This has led some works~\cite{Erlich:2006hq} to erroneously conclude that the Pad\'e approximant to the spectrum of QCD arises naturally in 5-dimensional models with AdS metric. However, after the discussion of the last section, this statement cannot be right, as first pointed out in~\cite{Cata:2006ak}. 

The reasons why both approaches yield the same results have been discussed in detail in~\cite{Falkowski:2006uy}. Roughly speaking, it has to be attributed to three common ingredients, which are implemented in very different ways:
\begin{itemize}
\item {\bf{conformal invariance}}. The Lagrangian of QCD in the massless limit is scale invariant. Thus, conformal invariance is implemented by hand in Migdal's construction by choosing the leading perturbative logarithm as input. In the holographic model the presence of the logarithm is just an automatic consequence of conformal invariance of the AdS metric.
\item {\bf{infinite spectrum}}. Through meromorphization in Migdal's construction, and by compactification in the holographic model. Actually, the 5-dimensional equations of motion for the $V_{\mu}$ field can be cast as a quantum mechanical eigenvalue problem.
\item {\bf{confinement scale}}. Introduced in Migdal's construction through a sophisticated continuum limit, and set by the infrared brane in the holographic model.   
\end{itemize}
Albeit at present both approaches are rather {\emph{ad hoc}} mechanisms to mimic confinement, the similarities between them are extremely interesting. Applications of the AdS/CFT correspondence to QCD require a departure of the conformal limit in order to take confinement into account. If one knew how to implement this in a solid way, one would have valuable insight on how the string dual of QCD looks like, one of the holy grails in the string community.  If progress can be made on the 4-dimensional side with Pad\'e approximants (perhaps along the lines discussed at the end of the last section), then the gauge/string duality could be used to make contact with string theory constructions.

\subsection{Pad\'e approximants in Euclidean space: relations between low and high energy parameters}\label{sec43}
I will finally report on a different application of Pad\'e approximants, namely the prediction of high energies from known low energies. For the sake of illustration, I will concentrate on the $\Pi_{LR}$ correlator, with generic meromorphic ans\"atze of the form
\begin{equation}\label{correl}
\Pi_{LR}(q^2)=\frac{f_{\pi}^2}{q^2}+\sum_n^{{\cal{N}}_V}\frac{f_{Vn}^2}{-q^2+m_{Vn}^2}-\sum_n^{{\cal{N}}_A}\frac{f_{An}^2}{-q^2+m_{An}^2}~,
\end{equation}
The discussion will follow closely Ref.~\cite{Cata:2009fd}.

I already mentioned that at high energies, the first two terms of the OPE vanish and therefore the correlator converges like $q^{-6}$, meaning that $\xi_2=0$ and $\xi_4=0$. A determination of the leading $\xi_6$ and subleading $\xi_8$ coefficients of the OPE from experiment turns out to be problematic, and different analyses find serious discrepancies (see the first two columns of Table~1, where existing phenomenological determinations are summarized). The only solid aid from theoretical considerations comes from a theorem by Witten~\cite{Witten:1983ut}, which can be cast as the inequality
\begin{equation}
q^2\Pi_{LR}(q^2)\geq 0, \quad -\infty\leq q^2\leq 0~.
\end{equation}  
The previous equation in particular implies that $\xi_6>0$, but nothing is known about $\xi_8$. Actually, notice from Table~1 that even its sign is disputed.
 
However, it turns out that rather accurate experimental data exists on ${\mathrm{Im}}\,\Pi_{LR}(t)$ at low energies and in principle the chiral parameters (the MacLaurin coefficients) of $\Pi_{LR}$ can be determined with reliability. Taking $q^2\Pi_{LR}(q^2)$ as given by Eq.~(\ref{correl}) as a Pad\'e approximant, relations between high and low energy parameters can be established. The minimal version of Eq.~(\ref{correl}) consists of just one vector and one axial contribution, subject to the following constraints:
\begin{eqnarray}\label{2PPA}
\xi_2\equiv-\int_0^{\infty}dt\,{\mathrm{Im}}\,\Pi_{LR}(t)&=&
f_{A}^2-f_{V}^2+f_{\pi}^2=0~,\nonumber\\
\xi_4\equiv-\int_0^{\infty}dt\,t\,{\mathrm{Im}}\,\Pi_{LR}(t)&=&f_{A}^2m_{A}^2-f_{V}^2m_{V}^2=0~,\nonumber\\
\zeta_2\equiv\int_0^{\infty}\frac{dt}{t}{\mathrm{Im}}\,\Pi_{LR}(t)&=&\frac{f_{V}^2}{m_{V}^2}-\frac{f_{A}^2}{m_{A}^2}~,\nonumber\\
\zeta_4\equiv\int_0^{\infty}\frac{dt}{t^2}{\mathrm{Im}}\,\Pi_{LR}(t)&=&
\frac{f_{V}^2}{m_{V}^4}-\frac{f_{A}^2}{m_{A}^4}~,
\end{eqnarray}
where the first two equations are the high-energy constraints and the last two are the low energy matching equations coming from the MacLaurin expansion (chiral expansion) of the correlator, defined as
\begin{equation}
\lim_{q^2\rightarrow 0}\Pi_{LR}(q^2)=\frac{f_{\pi}^2}{q^2}+\sum_{j}\zeta_{2j}q^{2j}~.
\end{equation}
The parameters $f_V$, $f_A$, $m_V$ and $m_A$ are therefore determined from Eqs.~(\ref{2PPA}) as functions of $f_{\pi}$, $\zeta_1$ and $\zeta_2$. The expressions for $\xi_6$ and $\xi_8$ turn out to be rather simple~\cite{Cata:2009fd}:
\begin{eqnarray}
\xi_6&=&f_{A}^2 m_{A}^4-f_{V}^2 m_{V}^4\,\,\,=\,\,\,\frac{f_{\pi}^6}{\zeta_1^2-\zeta_2f_{\pi}^2}~,\label{cor1}\\
\xi_8&=&f_{A}^2 m_{A}^6-f_{V}^2 m_{V}^6\,\,\,=\,\,\,\frac{\zeta_1f_{\pi}^8}{(\zeta_1^2-\zeta_2f_{\pi}^2)^2}~.\label{cor2}
\end{eqnarray}
There are a set of consequences that can be readily inferred from the previous equations:
\begin{itemize}
\item Since it is experimentally established that $\zeta_1>0$, Eq.~(\ref{cor2}) immediately implies that $\xi_8>0$. Therefore, the $P^{0}_{2}$ Pad\'e approximant we are considering here favours the set of phenomenological analyses sitting on the first half of Table~1.
\item Combining Eqs.~(\ref{cor1}) and (\ref{cor2}) one can get the relation
\begin{equation}\label{rel}
\xi_8=\frac{\zeta_1}{f_{\pi}^4}\xi_6^2~,
\end{equation}
which turns out to be fulfilled by the same first half of Table~1 to a remarkable degree of accuracy, as shown in Fig.~\ref{fig5}. Therefore, even though the different phenomenological determinations differ on the values of $\xi_6$ and $\xi_8$, it seems as if Eq.~(\ref{rel}) is a universal constraint on those determinations with $\xi_8>0$. 
\item Witten's inequality on Eq.~(\ref{cor1}) sets an upper bound on $\zeta_2$:
\begin{equation}
\zeta_2<\left(\frac{\zeta_1}{f_{\pi}}\right)^2~,
\end{equation}
which can be used as a consistency check of the full approach. Plugging typical numbers for $\zeta_1$ and $f_{\pi}$, one gets $\zeta_2< 0.1$ GeV$^{-2}$, which agrees well with the typical $\zeta_2\simeq 0.08$ GeV$^{-2}$.
\end{itemize} 
\begin{table}[t]
\renewcommand{\arraystretch}{1.4}
\setlength{\doublerulesep}{0.15mm}
\begin{center}
\begin{tabular}{cccc}
\hline\hline
 &  $\xi_6$  & $\xi_8$  & $\xi_8=\zeta_1f_{\pi}^{-4}\xi_6^2$\\
\hline 
\hline
Friot {\it{et al.}}~\cite{Friot:2004ba} & $+7.90\pm 1.63$ \,\,\,\,\,\, & $+11.69\pm 2.55$  \,\,\,\,\,\, & $+9.0\pm 3.7$ \\
\hline
Ioffe {\it{et al.}}~\cite{Ioffe:2000ns}  & $+6.8\pm 2.1$ & $+7\pm 4$  & $+6.7\pm 4.1$ \\
\hline
Zyablyuk~\cite{Zyablyuk:2004iu}  & $+7.2\pm 1.2$ & $+7.8\pm 2.5$  &  $+7.5\pm 2.5$\\
\hline
Narison~\cite{Narison:2004vz}  & $+8.7\pm 2.3$ & $+15.6\pm 4.0$  &  $+10.9\pm 5.8$ \\
\hline
ALEPH~\cite{Davier:2005xq}  & $+8.2\pm 0.4$ & $+11.0\pm 0.4$  &   $+9.71\pm 0.96$\\
\hline
OPAL~\cite{Ackerstaff:1998yj}  & $+6.0\pm 0.6$ & $+7.6\pm 1.5$  &  $+5.2\pm 1.0$ \\
\hline
\hline
Cirigliano {\it{et al.}} on ALEPH~\cite{Cirigliano:2003kc} \,\,\, & $+4.45\pm 0.70$ & $-6.16\pm 3.11$  &  $+2.86\pm 0.90$ \\
\hline
Cirigliano {\it{et al.}} on OPAL~\cite{Cirigliano:2003kc} \,\,\, & $+5.43\pm 0.76$ & $-1.35\pm 3.47$  &  $+4.3\pm 1.2$ \\
\hline
Bijnens {\it{et al.}} on ALEPH~\cite{Bijnens:2001ps} & $+3.4^{+2.4}_{-2.0}$ & $-14.4^{+10.4}_{-8.0}$  &   $+1.7\pm 2.4$\\
\hline
Bijnens {\it{et al.}} on OPAL~\cite{Bijnens:2001ps} & $+4.0\pm2.0$ & $-10.4^{+8.0}_{-6.4}$  &  $+2.3\pm 2.3$ \\
\hline 
Latorre {\it{et al.}}~\cite{Rojo:2004iq} & $+4.0\pm2.0$ & $-12^{+7}_{-11}$  &   $+2.3\pm 2.3$\\
\hline 
Almasy {\it{et al.}}~\cite{Almasy:2008xu}  & $+3.2^{+1.6}_{-0.4}$ & $-17.0^{+2.5}_{-9.5}$  &  $+1.5\pm 1.5$\\
\hline
\hline
{\bf{Ref.~\cite{Cata:2009fd}}}& $+7.6\pm0.4$ & $+8.3\pm1.0$  & $$ \\
\hline
\hline
\end{tabular}
\end{center}
\caption{Values for the dimension-six and dimension-eight OPE condensates (in units $10^{-3}$ GeV$^6$ and $10^{-3}$ GeV$^8$, respectively) reported using different phenomenological techniques. In the last column I list the would-be value for the dimension-eight condensate if Eq.~(\ref{rel}) were used, taking as inputs typical values for $\zeta_1$ and $f_{\pi}$ and the different values of $\xi_6$.\label{tab1}}
\end{table} 
The exercise considered before is just the first step of an iterative process, where one should consider the sequence of Pad\'e approximants $P^{N}_{N+2}$ subject to an increasing number of low energy input in the form of the MacLaurin coefficients $\zeta_{2j}$. This will give rise to an associated sequence of predictions for $\xi_6$ and $\xi_8$. What is remarkable is the non-trivial agreement with the first half of Table~1, which suggests that already $P^{0}_{2}$ might yield values for $\xi_6$ and $\xi_8$ close to the real ones. Should this be true, then we would have a criteria that strongly favours $\xi_8>0$.

This kind of strategy is especially suited for problems where only Euclidean quantities, {\it{i.e.}}, those defined or involving only the Euclidean half-plane, are requested. Notice however that this approach differs from the MHA introduced in Section~\ref{sec32}. The MHA is an exercise in Pad\'e interpolation, while the approach discussed here is an exercise in Pad\'e extrapolation. In fact, extrapolation from $q^2=0$ to arbitrary large values of $q^2$.\footnote{Notice that the problem falls out of the scope of Pommerenke's theorem~\cite{Pommerenke} because the OPE is defined at $q^2\rightarrow \infty$, {\it{i.e.}}, not in a compact subset around the origin.} Notice also that I use Pad\'e approximants instead of Pad\'e-type ones.
\begin{figure}[t]
\begin{center}
\includegraphics[width=8cm]{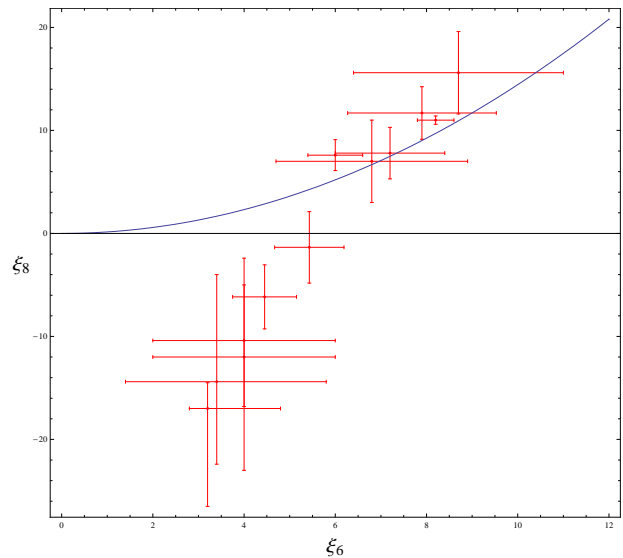}
\caption{Values of the OPE condensates from the different phenomenological analyses and their comparison with Eq.~(\ref{rel}).}\label{fig5}
\end{center}
\end{figure}
 
\section{Conclusions}\label{sec5}
I have presented a brief overview of commonly used methods to compute non-perturbative quantities in QCD and their close relationship with Pad\'e approximants to meromorphic functions. All those methods were originally developed without prior knowledge of the developments in Pad\'e theory. Only recently this connection has been acknowledged and its far-reaching consequences are only starting to be fully developed~\cite{Masjuan:2007ay,Cata:2009fd}. It is quite remarkable how concepts like meromorphization come out rather naturally in QCD if one embraces the large-$N_c$ limit; quark-hadron duality guarantees the matching between pQCD (and the OPE) and the hadronic spectrum; or order parameters of chiral symmetry breaking avoid the presence of high energy logarithms. The fact that the different physical approaches ended up converging to well-established methods in numerical analysis reveals Pad\'e approximants as extremely useful tools to explore the non-perturbative aspects of QCD.

Certain problems, like the ones I presented in Sections~\ref{sec41}, \ref{sec42} and \ref{sec43}, fall out of the scope of such techniques and require new methods. I am convinced that strategies based on Pad\'e approximants can lead us far, but attempts have been scarce so far.

In writing this article I had to omit quite a lot of material, such as the applications of Pad\'e theory to heavy quark physics~\cite{Broadhurst:1993mw,Chetyrkin:1996cf,Hoang:2008qy,Masjuan:2009wy}, the interesting works on light quarks of Refs.~\cite{Masjuan:2008fr,Masjuan:2008fv}, Pad\'e unitarization~\cite{Masjuan:2008cp} or QCD with finite temperature and chemical potential~\cite{Lombardo:2005ks}. Hopefully the contributions of S.~Peris and J.~J.~Sanz-Cillero in this conference can help fill some of this gaps. 





\end{document}